\documentclass{aa}

\usepackage{psfig,natbib}

\begin{document}

%% abbreviations 

\newcommand{\crii}{Cr\,{\sc ii}}
\newcommand{\feii}{Fe\,{\sc ii}}
\newcommand{\feiii}{Fe\,{\sc iii}}
\newcommand{\ciii}{C\,{\sc iii}]}
\newcommand{\civ}{C\,{\sc iv}}
\newcommand{\mgii}{Mg\,{\sc ii}}
\newcommand{\znii}{Zn\,{\sc ii}}
\newcommand{\micron}{\mbox{$\mu$m}}
\newcommand{\tabsp}{\noalign{\smallskip}}
\newcommand{\kms}{\mbox{km~s$^{-1}$}}
\newcommand{\lalala}{$\lambda$$\lambda$$\lambda$}
\newcommand{\qo}{\mbox{q$_0$}}
\newcommand{\Hoa}{\mbox{H$_0= 75$~km~s$^{-1}$~Mpc$^{-1}$}}
\newcommand{\Lir}{\mbox{$L_{\mbox{\tiny IR}}$}}
\newcommand{\Lo}{\mbox{L$_{\odot}$}}
\newcommand{\x}{\mbox{$\times$}}
\def\rf@jnl#1{{#1}}
\def\apj{\rf@jnl{ApJ }}   
\def\mnras{\rf@jnl{MNRAS }} 
\def\apjl{\rf@jnl{ApJ }}
\def\aap{\rf@jnl{A\&A }}
\def\apjs{\rf@jnl{ApJS }}  
\def\aj{\rf@jnl{AJ }}

\title{An unusual  
iron Lo-BAL quasar detected by ISOCAM
\thanks{Based on observations with ISO, an ESA project with instruments funded by ESA Member States
with the participation of ISAS and NASA and observations  collected at the European Southern
Observatory, Paranal, Chile (ESO No 65.O-0563 A), at La Silla, Chile (ESO No 67.A-0253 A) 
 }}

   \author{P.--A. Duc \inst{1,2}  \and P. B. Hall \inst{3}
   \and  D. Fadda \inst{4}  \and   P. Chanial \inst{2} \and  D. Elbaz \inst{2}
   \and P. Monaco \inst{5} \and  E. Pompei \inst{6}  \and B.M. Poggianti \inst{7}
   \and  H. Flores  \inst{8} \and  A. Franceschini \inst{9} \and A. Biviano  \inst{10} 
   \and A. Moorwood \inst{11} \and  C. Cesarsky \inst{11}}

   \offprints{paduc@cea.fr}

   \institute{
CNRS URA 2052
\and
CEA, DSM, DAPNIA, Service d'astrophysique, 91191 Gif--sur--Yvette Cedex, France
\and
Pontificia Universidad Cat\'{o}lica de Chile, Departamento de
Astronom\'{\i}a y Astrof\'{\i}sica, Casilla 306, Santiago 22, Chile
and Princeton University Observatory, Princeton, NJ 08544-1001, USA
\and
Instituto de Astrofisica de Canarias, Via Lactea s/n, E--38200 La Laguna -- Tenerife, Spain
\and
Dipartimento di Astronomia, via G.B. Tiepolo 11, 34131 Trieste, Italy
\and
European Southern Observatory, Santiago
\and
Osservatorio Astronomico di Padova, vicolo dell'Osservatorio 5, 35122 Padova, Italy
\and
DAEC/LUL, Observatoire de  Paris--Meudon, 5 place Jules Janssen, 92195 Meudon, France
\and
Dipartimento di Astronomia, Universit\`a di Padova, Vicolo dell'Osservatorio, 5, I35122 Padova, Italy
\and
INAF -- Osservatorio Astronomico di Trieste, via G.B. Tiepolo, 11, I--34131 Trieste, Italy
\and 
European Southern Observatory, Karl--Schwarzschild--Strasse, 2, D--85748 Garching bei M\"unchen, Germany
}
   \date{Accepted as a letter to A\&A}

   \titlerunning{An unusual FeLoBAL quasar detected by ISOCAM}

   \authorrunning{P.--A.\ Duc et al.}

\abstract{
We report the discovery of an unusual low--ionization broad absorption line 
quasar at $z=1.776$ which exhibits  
absorption lines from many excited states of \feii.
This member of the rare class of 'FeLoBAL' quasars was serendipitously found in a 
mid--infrared (MIR) survey of distant clusters carried out with the ISOCAM camera.
ISO~J005645.1$-$273816 has a high MIR to UV luminosity ratio, suggesting a strong 
dust obscuration plus emission from very hot dust. This characteristic
 makes MIR surveys particularly efficient at detecting  LoBAL quasars.
}
\maketitle

%-----------------
\section{Introduction}
%-----------------
About 10\% of optically selected quasars show absorption from gas with
blueshifted outflow velocities of typically $\la$0.1$c$ \citep{Weymann91}.
These Broad Absorption Line (BAL) quasars may just be normal quasars seen along
a particular line of sight, such that 
most quasars 
have BAL outflows covering $\sim$10--30\% of the sky, with mass loss rates 
comparable to the quasar accretion rates ($\sim1$\, $M_{\odot}$\,yr$^{-1}$).
 It is also possible that BAL quasars represent a dust--enshrouded early phase
in the lives of most, if not all, quasars \citep{Becker00}.  In either case,
BAL outflows must be understood to understand quasars as a whole.

BAL quasars are divided into three observational
subtypes depending on what type of absorption is seen.
{\em HiBAL} quasars show absorption from high--ionization lines like \civ.
{\em LoBAL} quasars \citep{Voit93}
also show absorption from low--ionization lines like \mgii.
{\em FeLoBAL} quasars \citep{Becker97} are LoBAL quasars which also show
absorption from excited 
\feii\ or \feiii.
Populations of unusual BAL quasars with extremely strong or complex absorption
have recently been found
through followup of FIRST radio sources \citep{Becker97,Becker00,Menou01,Lacy02},
$z>4$ quasar candidates from the Digitized Palomar Sky Survey \citep{Djorgovski01},
and color--selected quasar candidates from the Sloan Digital Sky Survey
\citep{Hall02}.
These unusual BAL quasars suggest that the range of physical conditions present
in BAL outflows is larger than previously suspected.

In this paper we report the discovery of another such unusual BAL quasar with
the Infrared Space Observatory (ISO).
What little data exists on the mid- to far--IR properties of BAL quasars
suggests that BAL quasars, and especially LoBAL quasars, may be much more
common or have larger covering fractions than suggested by optical surveys.
 For example, the dusty LoBAL quasar Hawaii 167 was discovered in a near--IR
survey covering only 77 square arcmin \citep{Cowie94},
and $\sim20_{-10}^{+15}$\% of {\em all} IRAS--selected quasars are
LoBAL quasars \citep{Weymann91,Low89,Boroson92a} 
vs. only $\sim$1.5\% in optical surveys.
Dusty gas with a nearly 100\% covering factor in LoBAL quasars
helps explain many of their unusual properties, including
ubiquitous signs of recent mergers in their host galaxies \citep{Canalizo01}.

%-----------------
\section{Observations}
%-----------------
The quasar ISO~J005645.1$-$273816 (hereafter ISO~0056$-$2738) was serendipitously discovered
 as part of an on--going
study of distant clusters of galaxies mapped in the mid--infrared by the ISOCAM
camera on board the ISO satellite.
This quasar was found in the field of the $z=0.56$ cluster J1888.16CL.

J1888.16CL  was observed with ISOCAM at
6.75 and 15 $\mu$m in December 1997 and with ISOPHOT at 200 $\mu$m
in December 1996.
The ISOCAM data were processed following the method
presented by \citet{Fadda00} which uses extensive simulations
based of the addition of fake sources to the science images to estimate
flux reconstruction factors as well as error estimates.
ISOPHOT data were reduced with the standard pipeline PIA \citep{Gabriel97}.

B and R images of J1888.16CL  were  obtained in August and September 1999 with
the WFI camera installed on the MPG/ESO 2.2m telescope at La Silla, as part
 of a program on supernova monitoring (PI: Patat). Near--infrared J,H and Ks
 broad--band images of the field were
taken in August 2001 with SOFI on the NTT at La Silla.
We retrieved  a F814W image  from the HST archives.
The optical spectrum of ISO~0056$-$2738 was obtained in September 2000 with the FORS1
instrument installed on the VLT UT1/Antu. The medium--resolution grism  300V (resolution
420) covering the wavelength range 4200\AA -- 8820\AA\ was used associated
with a slitlet of 1.2\arcsec\ positioned with the MOS unit.

%-----------------
\section{Results}
%-----------------
\label{sec:results}

\subsection{Source identification}
\begin{figure}
\centerline{\psfig{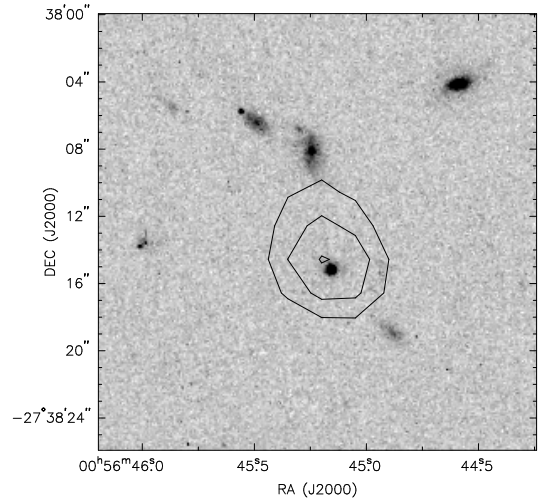}}
\caption{ISOCAM LW3 (15 $\micron$) contours  superimposed on a
combined F606W+F814W  HST image centered on the quasar. The levels are
2, 2.5 and 3 $\mu$Jy arcsec$^{-2}$.}
\label{fig:field}
\end{figure}

With a flux of 1.3 mJy, ISO~0056$-$2738 is one of the brightest 15\micron\ extragalactic
 sources in the field of J1888.16CL.  It is unambiguously associated with an object
which appears unresolved in the optical and near--infrared bands.
  On the HST image, its FWHM is lower than
0.22\arcsec, and no clear extension is visible (see Fig.~\ref{fig:field}).

The basic properties of the object, including position and photometry, are listed in
Table~\ref{tab:data}.
ISO~0056$-$2738 was not detected by IRAS at 60 and 100 $\micron$,
nor at 200~$\micron$ by ISOPHOT or at 20~cm  by
the  NRAO VLA Sky Survey (NVSS).

\begin{table}
\caption{Data}
\begin{tabular}{ll}
\hline \tabsp
RA (J2000)  & 00:56:45.15 \\
DEC (J2000) & -27:38:15.6 \\
\tabsp \hline \tabsp
B (0.46 $\micron$) & 22.74 $\pm$  0.03 mag \\
R (0.65 $\micron$) & 20.95 $\pm$  0.02 mag \\
J (1.25 $\micron$) & 18.29 $\pm$ 0.04 mag \\
H (1.65 $\micron$) & 17.68 $\pm$ 0.04 mag \\
Ks (2.16 $\micron$) & 17.16 $\pm$ 0.04 mag \\
LW2 (6.75 $\micron$) & 0.51 $\pm$ 0.15 mJy \\
LW3 (15 $\micron$) & 1.33 $\pm$ 0.33 mJy  \\
PHOT (200 $\micron$) &  $<$ 1 Jy \\
VLA NVSS (20~cm) &  $<$ 2.3 mJy \\
\tabsp \hline
\end{tabular}
\label{tab:data}
\end{table}

\subsection{Spectroscopic Analysis}  \label{SPEC}

\begin{figure*}
\centerline{\psfig{file=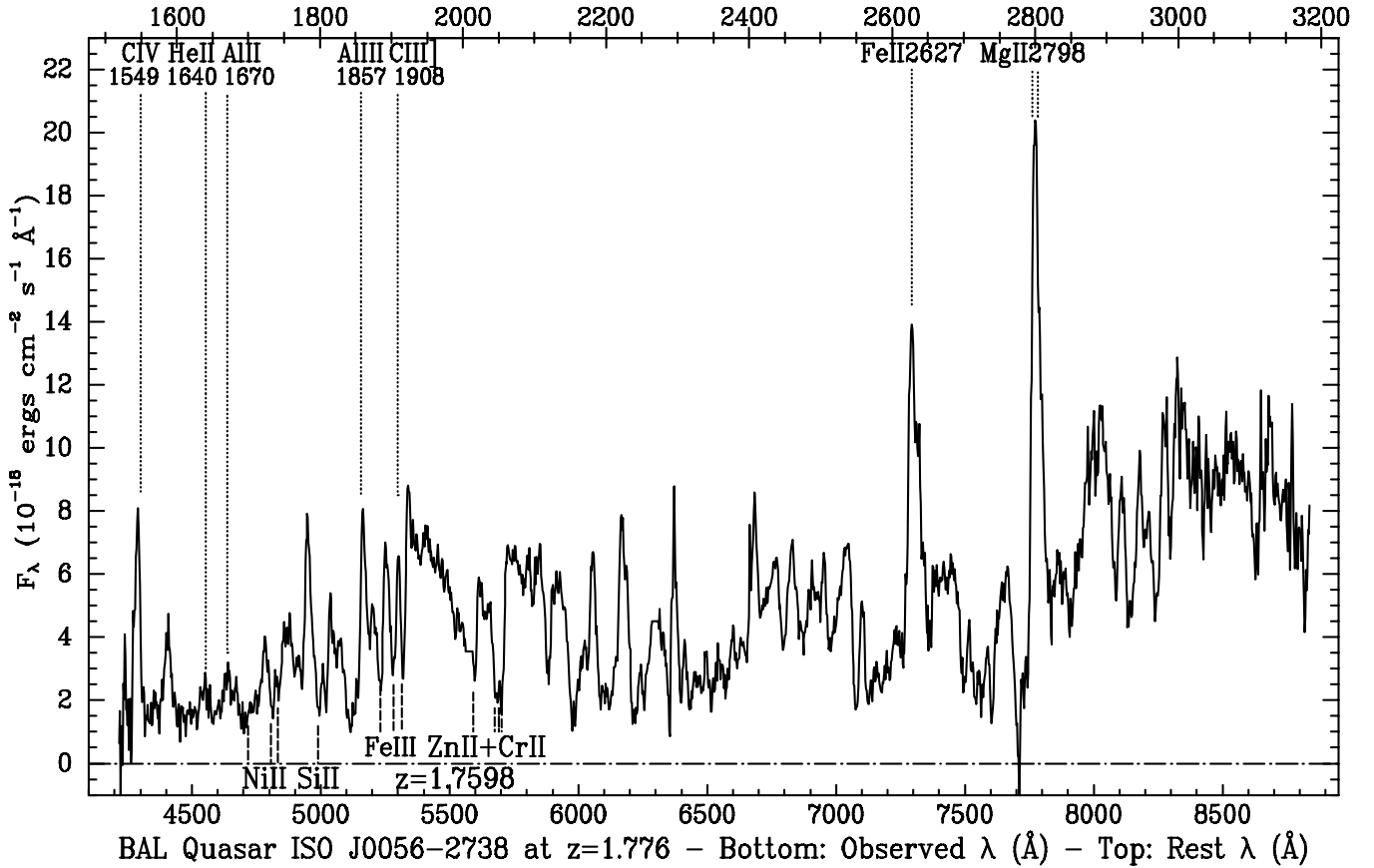,width=\textwidth}}
\caption{
The VLT spectrum of ISO~0056$-$2738.  
The horizontal dot--dashed line is the zero flux level.
Dotted lines above the spectrum
show the wavelengths of the strong emission lines labeled across
the top of the plot (including the expected wavelength of the \ciii\ line peak).
Dashed lines below the spectrum show the wavelengths of the strong absorption
lines from the $z=1.7598$ system labeled across the bottom of the plot.}
\label{fig:spec}
\end{figure*}

The VLT spectrum of  ISO~0056$-$2738 is shown in Figure~\ref{fig:spec}.
Extensive UV absorption plus narrow emission from \feii\,UV1 and \mgii\
identifies it as an unusual FeLoBAL quasar similar to
FIRST~0840+3633 \citep{Becker97} and especially PSS~1537+1227 \citep{Djorgovski01}.

We adopt a systemic redshift of $z=1.776\pm0.002$ based on the
narrow emission lines of \civ, \feii~2627 (multiplet UV1) and \mgii,
using the effective wavelengths for those transitions in the first SDSS
composite quasar \citep{VandenBerk01}.
As long as the relative velocity shifts of the lines in this quasar do not
differ greatly from those quasars used to construct the composite, this should
yield consistent redshifts for all lines.
Indeed, the three redshifts agree to better than $1 \sigma$.
Of course, since all three emission lines are affected by absorption, the true
systematic emission line redshift may lie blueward of our adopted value.
The balnicity index \citep[BI; ][]{Weymann91} of this object,
measured from Al\,{\sc iii}, is a small but nonzero 305\,\kms.
The less restrictive absorption index  \citep[AI; ][]{Hall02} is 2860\,\kms.
Both measures assume the Al\,{\sc iii} trough is at least 6280\,\kms\ wide,
though it could be confused with a different trough beyond 4000\,\kms.

 At least two different redshift systems are present in the absorption.
The \civ\ and \mgii\ troughs appear to reach zero flux at $1.7487\pm0.0008$.
The narrow absorption from the \feiii\ UV34 multiplet
(\lalala1895.46,1914.06,1926.30) gives $z=1.7598\pm0.0009$.
There are no obvious features in the \civ\ or \mgii\ troughs at this redshift,
but there is absorption from
Ni\,{\sc ii}\,\lalala1709,1741,1751, Si\,{\sc ii}\,$\lambda$1808,
and \znii\,$\lambda\lambda$2026,2062 plus \crii\,\lalala2056,2062,2066.

Overall, however, the spectrum is dominated by absorption
from ground and excited terms of \feii.  Essentially every absorption trough
not labelled in Figure~\ref{fig:spec} can be identified with \feii.
For example, the strong absorption just redward of \civ\ at 1550--1600\,\AA\ is
from \feii\ UV multiplets 44--46 \citep{Wampler95,Hall02}.
\feii\ multiplets from excited atomic terms are present up to at least
UV191 ($\sim$1787\,\AA, excitation potential $\sim$2.88\,eV).  There may also
be excited \crii\ UV5--UV8 absorption near 2675 and 2835\,\AA\ \citep{deKool02}. 
 The 3130\,\AA\ trough could have contributions from even more highly
excited \crii\ and possibly O\,{\sc iii}\,$\lambda$3133, as discussed in
section 5.1.3 of \citet{Hall02}.

Given the strength of the \feii\ absorption, the \mgii\ absorption must be
saturated.  However, the flux in the absorption troughs does not reach zero
except for narrow regions in the \civ\ and \mgii\ troughs.  This means that
the absorbing region does not cover the continuum source except at a narrow
range of velocities, and that the \feii\ absorbing region probably
has a smaller partial covering factor than the \mgii\ and \civ\ regions.

\section{Discussion and conclusions}
\begin{figure}
\centerline{\psfig{file=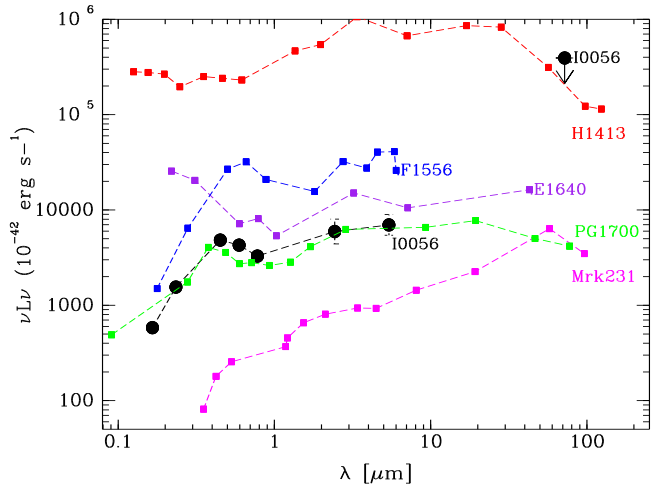,width=9cm,angle=-90}}
\caption{
Rest--frame spectral energy distribution of ISO~J005645.1$-$273816 (plain circles) and, for  comparison, of
other BAL quasars with IR data: Mrk~231 (LoBAL; z=0.04;data from NED); 
PG~1700+518 \citep[LoBAL; z=0.292; ][]{Andreani99},  
ELAIS~J164010+410502 \citep[BAL; z=1.099; ][]{Morel01};
FIRST~J155633.7+351757 \citep[LoBAL; z=1.48; ][]{Clavel98}, 
H1413+117 \citep[BAL; z=2.54; ][]{Barvainis95}. Note the
lines joining the data points, indicated by plain squares, were drawn to guide
the eye and by no means represent the real SEDs. We used \Hoa\ and \qo=0.5.
}
\label{fig:sed}
\end{figure}

The spectral energy distribution of  ISO~0056$-$2738 in the rest--frame UV to infrared
domain is compared in Figure~\ref{fig:sed} to that of a sample of BAL quasars for
which  mid and far infrared  fluxes are available in the literature.

The SED of ISO~0056$-$2738 appears to be  similar, in shape and luminosity,
to that of the nearby optically selected low ionization BAL quasar  PG 1700+518, at least
within the wavelength range for which data are available for both
objects.
If this similarity extends to the far infrared where IRAS fluxes are available for PG 1700+518,
 one would infer for ISO~0056$-$2738 a total infrared luminosity of $\Lir = 4 \x 10^{12}~\Lo$.
Instead one may use as template Mrk 231, which is known as the nearest
 galaxy with a LoBAL signature, and has a strong IR bump at 60  $\micron$.
Taking the rest--frame 5 $\micron$ luminosity as the normalizing factor,
 the  total infrared luminosity of ISO~0056$-$2738 becomes $2 \x 10^{13}~\Lo$.
The PHOT measurement at 200  $\micron$ flux indicates that this is
 most likely an upper limit.
One may conclude for this analysis that unless its SED
plunges at $\lambda > 10 \micron$, ISO~0056$-$2738 belongs to the class of ULIRGs and
may even be a hyperluminous infrared galaxy,  like many quasars
\citep[e.g.][]{Haas00} .
ISO~0056$-$2738 lies behind a rich cluster of galaxies; however the distance
from the cluster center, about $3.5'$, is such that a gravitational lensing
by the cluster is unlikely.

The spectral energy distribution of the quasar in the rest--frame ultraviolet
is much redder than that of the SDSS composite quasar of \citet{VandenBerk01}.
Using the SMC extinction curve of \citet{Prevot84}, we estimate a
large but not unprecedented extinction of $E(B-V)=0.25\pm0.05$.  
Besides, ISO~0056$-$2738 seems to share with  the few other LoBAL quasars with
 available mid--IR data (see Fig.~\ref{fig:sed}) a steep UV to MIR spectral index.
In addition to the extinction in the UV, its very high MIR to UV luminosity ratio
(rest--frame $\nu L_{\nu}$(5.40$\micron$)/ $\nu L_{\nu}$(0.16$\micron$) $=12$ )
 is likely due to the presence of unusually large quantities of very hot dust
at 1000 K, the emission of which peaks at about 3$\micron$.
 \citet{Becker97} claim that a relative excess of radio emission could
 be another property of LoBAL quasars. Note that ISO~0056$-$2738 is 
not likely to be radio--loud (log($R^*$)$>$1).
It has log($R^*$)$<$1.79 using the \citet{Becker00} definition
with $\alpha_{\rm rad}=-0.5$, but corrected for reddening of $E(B-V)=0.25$
it has intrinsic log($R^*$)$<$1.05.

To our knowledge, ISO~0056$-$2738 is the first FeLoBAL quasar so far directly found at
mid--infrared wavelengths  \footnote{However several  BAL quasars were identified in
the course of  the 15$\micron$ European Large Area ISO Survey (ELAIS) 
  \citep[e.g.][]{Morel01,Alexander01} }.
   At optical wavelengths,
the SDSS Early Data Release Quasar Sample \citep{Schneider02} only contains
four FeLoBAL quasars out of 200--300 BAL quasars at $1.485 < z < 3.9$
where both HiBAL and LoBAL quasars can be selected via SDSS spectra,
and three or four more FeLoBAL quasars at lower redshift \citep{Hall02}. Therefore
the probability of a serendipitous discovery with ISOCAM of such
 FeLoBAL quasars is then extremely low unless  mid--IR surveys are very efficient
 in detecting such objects.
 A quasar like ISO~0056$-$2738
 would likely have been missed by most optical
 surveys since it does not show any  prominent broad emission lines and has no UV excess.
 The  infrared surveys with SIRTF such as GOODS 
and  SWIRE  
 should soon confirm whether the number of FeLoBAL quasars  has
been underestimated in optically--based surveys.

\begin{acknowledgements}
PBH acknowledges financial support from Chilean grant FONDECYT/1010981.
 We are grateful to our referee, M. Lacy, for his comments which helped clarifying
the paper. 
We thank G. Rodighiero for her help in reducing the PHOT data, E. Cappellaro and F. Patat
 for providing us the optical images  of J1888.16CL. 
\end{acknowledgements}

%--------------------------------------------------------------------
%           Bibliography  (generated by bibtex)
%--------------------------------------------------------------------

\bibliographystyle{mybst}
%\bibpunct{(}{)}{,}{a}{}{,}
\bibliography{all}

\end{document}